\documentstyle[12pt]{article}
\catcode`@=11
\newcount\@tempcntc
\def\@citex[#1]#2{\if@filesw\immediate\write\@auxout{\string\citation{#2}}\fi
  \@tempcnta\z@\@tempcntb\m@ne\def\@citea{}\@cite{\@for\@citeb:=#2\do
    {\@ifundefined
       {b@\@citeb}{\@citeo\@tempcntb\m@ne\@citea\def\@citea{,}{\bf ?}\@warning
       {Citation `\@citeb' on page \thepage \space undefined}}%
    {\setbox\z@\hbox{\global\@tempcntc0\csname b@\@citeb\endcsname\relax}%
     \ifnum\@tempcntc=\z@ \@citeo\@tempcntb\m@ne
       \@citea\def\@citea{,}\hbox{\csname b@\@citeb\endcsname}%
     \else
      \advance\@tempcntb\@ne
      \ifnum\@tempcntb=\@tempcntc
      \else\advance\@tempcntb\m@ne\@citeo
      \@tempcnta\@tempcntc\@tempcntb\@tempcntc\fi\fi}}\@citeo}{#1}}
\def\@citeo{\ifnum\@tempcnta>\@tempcntb\else\@citea\def\@citea{,}%
  \ifnum\@tempcnta=\@tempcntb\the\@tempcnta\else
   {\advance\@tempcnta\@ne\ifnum\@tempcnta=\@tempcntb \else \def\@citea{--}\fi
    \advance\@tempcnta\m@ne\the\@tempcnta\@citea\the\@tempcntb}\fi\fi}
\catcode`@=12
\def\barr{\begin{array}}
\def\earr{\end{array}}
\def\beq{\begin{equation}}
\def\eeq{\end{equation}}
\def\bea{\begin{eqnarray}}
\def\eea{\end{eqnarray}}
\def\bmath{\begin{displaymath}}
\def\emath{\end{displaymath}}
\def\bq{\begin{quote}}
\def\eq{\end{quote}}
\def\slash#1{\setbox0=\hbox{$#1$}#1\hskip-\wd0\hbox to\wd0{\hss\sl/\/\hss}}
\begin{document}

\begin{flushright}
RAL/93-082 \\
October 1993
\end{flushright}

\begin{center}
{\bf{\large {\em CP} VIOLATION IN HIGGS DECAYS }}\\
{\bf{\large  DUE TO MAJORANA NEUTRINOS
\footnote[1]{To appear in
{\it Proceedings of the Workshop on $e^+e^-$ Collisions at 500~GeV:
The Physics Potential},
Munich, Annecy, Hamburg, 20 November 1992--3 April 1993,
ed.\ by P.M.~Zerwas.} }}\\[.5cm]
{\sf A. Ilakovac},$\!^{a}$
\footnote[2]{On leave from Department of Physics, University of Zagreb,
41001 Zagreb, Croatia}
{\sf B.A.~Kniehl},$\!^{b}$
{\sf and A.~Pilaftsis}$\,^{c}$
\footnote[3]{E-mail address: pilaftsis@vax2.rutherford.ac.uk}\\
$^{a}$ Institut f\"ur Physik, Johannes-Gutenberg Universit\"at,
55099 Mainz, Germany\\
$^{b}$ II.~Institut f\"ur Theoretische Physik, Universit\"at Hamburg,
22761 Hamburg, Germany\\
$^{c}$Rutherford Appleton Laboratory, Chilton, Didcot, Oxon, OX11 0QX, England
\end{center}

\vspace*{-1.cm}

\begin{center}
\parbox{14cm}
{\begin{center} {\bf Abstract} \end{center}
\vspace*{-0.3cm}

The possibility of observing $CP$ violation in the decays of Higgs particles
into top-quark, $W$- and $Z$-boson pairs induced by heavy Majorana neutrinos
is discussed. In the context of minimal ``see-saw" models with interfamily
mixings, we find that Majorana neutrinos may give rise to sizable $CP$-odd
effects at the one-loop electroweak order. We present numerical
estimates of typical $CP$-odd observables that might be triggered at
high-energy $e^+e^-$ colliders.}

\end{center}


There has been growing effort in the literature to analyze {\em gauge models}
and possible
mechanisms of producing sizable $CP$-violating effects at high-energy
colliders~\cite{CPhig,Peskin,CH,PN}.
Several scenarios that could account for
$CP$-violating effects at supercollider energies were considered. Most of
them~\cite{CPhig,Peskin,CH,PN}
were concerned with extensions of the Higgs sector of the
Standard Model ($SM$).
Here, we shall present the attractive possibility that heavy Majorana neutrinos
could play a crucial r\^ole in accounting for sizable $CP$-odd effects
at future $e^+e^-$ linear colliders~\cite{IKP}.

In the following,
we shall address the question whether loop graphs mediated by heavy
Majorana neutrinos can induce measurable $CP$-odd signals
in the decays $H^0\to t\bar{t}$, $H^0 \to W^+W^-$, and $H^0\to ZZ$.
In particular, we shall consider the following $CP$-odd asymmetry
parameters~\cite{Peskin,CH,PN}:
\bea
A_{CP}^{(t)} & = & \frac{\Gamma(H^0 \to t_L \bar{t}_L)\ -\
\Gamma(H^0 \to t_R \bar{t}_R) }{\Gamma(H^0 \to t \bar{t}\,) },\\
A_{CP}^{(W)} & = & \frac{\Gamma(H^0 \to W^+_{(+1)}W^-_{(+1)})\ -\
\Gamma(H^0 \to W^+_{(-1)}W^-_{(-1)})}
{\Gamma(H^0 \to W^+_{(+1)}W^-_{(+1)})\ +\ \Gamma(H^0 \to
W^+_{(-1)}W^-_{(-1)})},
\eea 
where the subscripts $L,R$, and $(\pm 1)$
denote the top-quark helicity and the $W$-boson transverse polarization,
respectively.
Similarly to Eq.~(2), a $CP$ asymmetry, $A_{CP}^{(Z)}$, for the decay
channel $H^0 \to ZZ$ can be constructed.
The states $t_L$ and $W^-_{(+1)}$ are connected with the states $\bar{t}_R$
and $W^+_{(-1)}$ by $CP$ conjugation.
Therefore, $A_{CP}^{(t)}$ and $A_{CP}^{(W)}$ represent
genuine $CP$-violating parameters that can be determined experimentally.
One should also remark that the $CP$ asymmetries $A_{CP}^{(t)}$,
$A_{CP}^{(W)}$,
and $A_{CP}^{(Z)}$ resulting from the Feynman graphs depicted in Fig.~1 cannot
be induced if the heavy neutrinos are of the standard Dirac type.\\[5cm]
\newcounter{fig}
{\small
\begin{list}{\bf\rm Fig. \arabic{fig}: }{\usecounter{fig}
\labelwidth1.6cm \leftmargin2.5cm \labelsep0.4cm \itemsep0ex plus0.2ex }

\item Feynman graphs giving rise to a $CP$-odd part in the
$H^0t\bar{t}$, $H^0W^+W^-$, and\\
$H^0ZZ$ couplings.

\end{list}
}
The minimal class of models that predict heavy Majorana neutrinos can be
obtained by simply introducing $n_G$ right-handed neutrinos, $\nu^0_{R_i}$,
to the $SM$, where $n_G$ denotes the number of generations.
The relevant interactions of the Majorana
neutrinos with the $W$-, $Z$-, and Higgs bosons in such minimal models
are given by~\cite{ZPC}
\bea
{\cal L}_{int}^W &=& -\ \frac{g_W}{2\sqrt{2}} W^{-\mu}\
{\bar{l}}_i \ B_{l_ij} {\gamma}_{\mu} (1-{\gamma}_5) \ n_j \quad + \quad H.c.\
,
\\[0.3cm]
{\cal L}_{int}^Z &=& -\ \frac{g_W}{4\cos\theta_W}  Z^\mu\
\bar{n}_i \gamma_\mu \Big[ i\mbox{Im}C_{ij}\ -\ \gamma_5\mbox{Re}C_{ij}
\Big] n_j\ ,\\[0.3cm]
{\cal L}^H_{int} &=& -\ \frac{g_W}{4M_W}\
H^0\ \bar{n}_i \Bigg[ (m_{n_i}+m_{n_j})\mbox{Re}C_{ij}
+\ i\gamma_5 (m_{n_j}-m_{n_i})\mbox{Im}C_{ij} \Bigg] n_j\ ,
\eea 
where
\bea
C_{ij}\ &=&\ \sum\limits_{k=1}^{n_G}\ U^\nu_{ki}U^{\nu\ast}_{kj}\ ,\\
B_{l_ij}\ &=& \sum\limits_{k=1}^{n_G} V^l_{l_ik} U^{\nu\ast}_{kj}\ .
\eea 
The matrices $B$ and $C$ satisfy a great number of identities that
will also help us to estimate our $CP$ asymmetries. These identities
are written down as~\cite{KPS}
\bea
\sum\limits_{i=1}^{2n_G} B_{l_1i}B_{l_2i}^{\ast} & = & \ {\delta}_{l_1l_2},\\
\sum\limits_{k=1}^{2n_G} C_{ik}C^\ast_{jk} & = & \ C_{ij}, \\
\sum\limits_{i=1}^{2n_G} B_{li}C_{ij} & = & \ B_{lj}, \\
\sum\limits_{k=1}^{2n_G} m_{n_i} C_{ik}C_{jk} & = & \ 0, \\
\sum\limits_{k=1}^{n_G} B_{l_ki}^{\ast}B_{l_kj} & = & \ C_{ij}.
\eea
The mixings $B_{lN}$ and $C_{\nu N}$ are constrained by a global analysis of
low-energy and LEP data~\cite{LL}. Notice that
Eqs.~(8)--(12) are not accidental,
but represent generalized GIM-type unitary relations
and guarantee the renormalizability of our model.

Note that the Lagrangians of Eqs.~(3)--(6)
violate the $CP$ symmetry of the model. Apart from the $CP$ violation
introduced by the
known complex Cabbibo-Kobayashi-Maskawa-type matrix, $B_{l_in_j}$,
the neutral-current interactions described by  Eqs.~(4) and~(5) violate
the $CP$ symmetry of the model, as the neutral particles, $H^0$ and $Z$,
couple simultaneously to $CP$-even ($\bar{n}_in_j$ or $\bar{n}_i\gamma_\mu
n_j$) and $CP$-odd
($\bar{n}_i\gamma_5n_j$ or $\bar{n}_i\gamma_\mu\gamma_5 n_j$) operators with
two different Majorana neutrinos (i.e., $n_i\neq n_j$).
As a result, the $H^0t\bar{t}$, $H^0W^+W^-$, and $H^0ZZ$ couplings
contain non-zero $CP$-odd parts which are generated radiatively
(see also Fig.~1)~\cite{IKP}.

We shall now look for $CP$-odd correlations of the
type $\langle(\vec{s}_t - \vec{s}_{\bar{t}} )\cdot\vec{k}_t\rangle$
in the decay $H^0\to t\bar{t}$.
The important ingredient for a non-zero $A_{CP}^{(t)}$ are Eqs.~(4) and (5),
which violate the $CP$ symmetry of the model, as mentioned above.
Thus, vacuum-polarization transitions between the $CP$-even Higgs and
the $CP$-odd $Z$ boson, which are not allowed in the $SM$, can now
give rise to a pseudoscalar part in the $H^0t\bar{t}$ coupling (see
also Fig.~1(a)) and hence lead to the non-vanishing $CP$ asymmetry
\beq
A_{CP}^{(t)}\ =\ \frac{\alpha_W}{4}\ \mbox{Im}C^2_{ij}
\sqrt{\lambda_i\lambda_j}
\ \frac{\lambda_i-\lambda_j}{\lambda_H}\
\frac{\lambda^{1/2}(\lambda_H,\lambda_i,\lambda_j)}
{\lambda^{1/2}(\lambda_H,\lambda_t,\lambda_t)}\ ,
\eeq 
where
\beq
\lambda_i \ = \ \frac{m_{n_i}^2}{M^2_W}, \quad \lambda_H\ =\
\frac{M^2_H}{M^2_W}, \quad \lambda_t\ =\ \frac{m^2_t}{M^2_W}, \quad
\lambda(x,y,z)\ =\ (x-y-z)^2-4yz.
\eeq 
To get $A_{CP}^{(t)}\neq 0$, two non-degenerate heavy Majorana neutrinos
are required. On the other hand, a counting of the
$CP$-odd phases existing generally in the $SM$ with right-handed
neutrinos gives~\cite{KPS}
\beq
{\cal N}_{CP}\ =\ N_L(N_R\ -\ 1),
\eeq 
where $N_L$ and $N_R$ are the numbers of left-handed and right-handed
neutrinos,
respectively. In the three-generation model, one has
Im$C^2_{N_1N_2} \leq 10^{-2}$~\cite{LL}. However, the situation changes
drastically if
one introduces an additional left-handed neutrino field, $\nu^0_{L_4}$, in the
Lagrangian of the $SM$~\cite{PH} and assumes a non-trivial mixing between
the two right-handed neutrinos.
Using Eq.~(14), one
can find the helpful relations
\bea
\mbox{Im}C^2_{\nu_4N_1}\ &=&\ \sin\delta_{CP}\ |C_{\nu_4N_1}|^2,\\
\mbox{Im}C^2_{\nu_4N_1}\ &=&\ -\frac{m_{N_1}}{m_{N_2}}\sin\delta_{CP}\
|C_{\nu_4N_1}|^2, \\
\mbox{Im}C^2_{N_1N_2}\ &=&\ \frac{m_{\nu_4}}{m_{N_2}}\sin\delta_{CP}\
|C_{\nu_4N_1}|^2.
\eea 
In these models, the mixing $|C_{\nu_4N_1}|^2$ can be of order one.
Detailed investigation reveals that, for $M_H\approx400$~GeV,
$CP$ asymmetries $A_{CP}^{(t)} \simeq 1.\ 10^{-2}$ are possible
in four-generation scenarios with Majorana neutrinos~\cite{IKP,PH}.
We again emphasize the fact that high-mass
Dirac neutrinos with standard diagonal couplings cannot
give rise to a non-zero $CP$ asymmetry.

Heavy Majorana neutrinos can also generate a $CP$-odd part
in the $H^0W^+W^-$ coupling through the triangle graph shown in Fig.~1(b).
Actually, one has to look
for $CP$-violating correlations of the form
\beq
\epsilon_{\mu\nu\rho\sigma}\ \varepsilon_{(+1)}^\mu(\hat{z})
\varepsilon_{(+1)}^\nu(-\hat{z}) k_+^\rho k_-^\sigma\ =
\ M_H \vec{k}_+\ (\vec{\varepsilon}_{(+1)}(\hat{z}) \ \times
\ \vec{\varepsilon}_{(+1)}(-\hat{z})),
\eeq 
where the polarization vectors,
$\varepsilon_{(+1)}^\mu(\hat{z})$ and $\varepsilon_{(+1)}^\nu(-\hat{z})$,
describe two transverse $W$ bosons with helicity $+1$. The presence of
$CP$-odd terms given in Eq.~(19) leads to the $CP$ asymmetry
\beq
A_{CP}^{(W)}\  =\ \frac{\alpha_W}{4} \Big[ \mbox{Im}(B_{l_ki}
B_{l_kj}^\ast C_{ij}^\ast)\sqrt{\lambda_i\lambda_j}\ (
F_+\ +\ F_- )\ +\ \mbox{Im}(B_{l_ki}B_{l_kj}^\ast C_{ij})
(\lambda_i F_+\ +\ \lambda_j F_-) \ \Big],
\eeq 
where $\alpha_W=(g_W^2/4\pi)$.
The functions $F_\pm$ in Eq.~(24) describe the absorptive contributions from
three different kinematic configurations of the intermediate states that
can become on-shell. A straightforward computation gives
\bea
F_\pm(\lambda_i,\lambda_j, \lambda_{l_k}) &=&
\theta(M_H-m_{n_i}-m_{n_j})\Bigg[ \ \pm\ \frac{\lambda^{1/2}
(\lambda_H,\lambda_i,\lambda_j)}{2\lambda^{1/2}(\lambda_H,\lambda_W,\lambda_W)}
\ +\ G_\pm(\lambda_i,\lambda_j,\lambda_{l_k})\nonumber\\
&& \times\ln\left( \frac{t^+(\lambda_i,\lambda_j)-\lambda_{l_k}}{
t^-(\lambda_i,\lambda_j)-\lambda_{l_k}}\right)\ \Bigg]\
+\ \theta(M_W-m_{n_i}-m_{l_k}) \Bigg[\Bigg(\ - \frac{\lambda_H-4\lambda_W}
{\lambda_W}\nonumber\\
&& \mp\ \frac{\lambda_H}{\lambda_W}\ \Bigg)\frac{\lambda^{1/2}(
\lambda_W,\lambda_i,\lambda_{l_k})}{4\lambda^{1/2}(\lambda_H,
\lambda_W,\lambda_W)}\ +\ G_\pm(\lambda_i,\lambda_j,\lambda_{l_k})
\ln\left( \frac{\bar{t}^+(\lambda_i,\lambda_{l_k})-\lambda_j}
{\bar{t}^-(\lambda_i,\lambda_{l_k})-\lambda_j}\right)\ \Bigg]\nonumber\\
&&+\ \theta(M_W-m_{n_j}-m_{l_k})\Bigg[\left(\ \frac{\lambda_H-4\lambda_W}
{\lambda_W}\ \mp\ \frac{\lambda_H}{\lambda_W}\ \right)\frac{\lambda^{1/2}(
\lambda_W,\lambda_j,\lambda_{l_k})}{4\lambda^{1/2}(\lambda_H,
\lambda_W,\lambda_W)}\nonumber\\
&&+\ G_\pm(\lambda_i,\lambda_j,\lambda_{l_k})
\ln\left( \frac{\bar{t}^+(\lambda_j,\lambda_{l_k})-\lambda_i}
{\bar{t}^-(\lambda_j,\lambda_{l_k})-\lambda_i}\right)\ \Bigg],
\eea 
where
\bea
\lambda_{l_k} &=& \frac{m_{l_k}^2}{M^2_W}, \qquad \lambda_W\ =\ 1,
\qquad \lambda_Z\ =\ \frac{M^2_Z}{M^2_W},\\
t^\pm(x,y) &=&
-\frac{1}{2}\Big[ \ \lambda_H-x-y-2\lambda_W
\ \mp\ \lambda^{1/2}(\lambda_H,x,y)
\lambda^{1/2}(\lambda_H,\lambda_W,\lambda_W)\  \Big],\\
\bar{t}^\pm(x,y)&=& \lambda_H+x - \frac{1}{2}\lambda_H(\lambda_W
+x-y)\ \pm \frac{1}{2}\lambda^{1/2}(\lambda_H,x,y)\lambda^{1/2}(\lambda_H,
\lambda_W,\lambda_W),\\
G_\pm(x,y,z)&=&\frac{x-y}{4\lambda_H}\ \pm\ \frac{2(\lambda_W+z)-
x-y}{4(\lambda_H-4\lambda_W)}.
\eea 
The dominant contribution to $A_{CP}^{(W)}$ comes from the $n_in_j$ on-shell
states.
To measure $A_{CP}^{(W)}$,
a discrimination between the events
$W^+_{(+1)}W^-_{(+1)}$ and $W^+_{(-1)}W^-_{(-1)}$ from the total number of
$W$ bosons produced by Higgs-boson decays is needed.
A branching-ratio estimate of these specific decay mode can be obtained from
the following ratio:
\bea
R^{(W)}\ &=&\ \frac{\Gamma(H^0 \to W^+_{(+1)}W^-_{(+1)})\ +\
\Gamma(H^0 \to W^+_{(-1)}W^-_{(-1)})}{\Gamma (H^0\to W^+W^-)}\nonumber\\
&\simeq & \frac{8M_W^4}{M^4_H}.
\eea 
Since one expects to produce 100--1000 Higgs bosons, with $M_H<400$~GeV,
at a 500-GeV $e^+e^-$ collider with an integrated luminosity of 20~fb$^{-1}$,
it should, in principle, be possible to see
$R^{(W)}$ values of order 1
and $CP$ asymmetries of $A_{CP}^{(W)} \simeq 10\%$
in the $SM$ with four generations.

\vspace*{-.3cm}

{\small
\newcounter{tab}
\begin{list}{\bf\rm Table \arabic{tab}: }{\usecounter{tab}
\labelwidth1.6cm \leftmargin2.5cm \labelsep0.4cm \itemsep0ex plus0.2ex }

\item Numerical results of $A^{(Z)}_{CP}/\mbox{Im}C^2_{N_1N_2}$ and $
A^{(Z)}_{CP}/\mbox{Im}C^2_{\nu_4N_1}$ for ``see-saw" models with\\
three and four generations, respectively.

\end{list}
}
\begin{tabular*}{6.62cm}{|r||cc|}
\hline
$m_{N_3}$ & $M_H$&$=\quad 400$~GeV \\
$[$GeV$]$& $m_{N_1}(m_{\nu_4})$&$=\quad 150$~GeV\\
& $m_{N_2}$&$=\quad 250$~GeV \\
 &$3Gens$&$4Gens$\\
\hline\hline
300 &-1.60~$10^{-2}$ &-1.41~$10^{-2}$ \\
400 &-3.70~$10^{-2}$ &-3.23~$10^{-2}$ \\
500 &-4.95~$10^{-2}$ &-4.29~$10^{-2}$ \\
600 &-5.74~$10^{-2}$ &-4.94~$10^{-2}$ \\
1000&-7.09~$10^{-2}$ &-6.02~$10^{-2}$ \\
\hline
\end{tabular*}

\vspace*{.3cm}

Finally, heavy Majorana neutrinos induce, through the triangle graph
of Fig.~1(c), a non-zero $CP$ asymmetry in the decay $H^0\to ZZ$,
which is computed to be
\bea
A_{CP}^{(Z)} \ &=&\ \frac{\alpha_W}{4}\ \Bigg[
\mbox{Im}(C_{ij}C_{jk}C_{ki})\ ( \lambda_iF_1\ +\ \lambda_jF_2 )
\ +\ \mbox{Im}(C_{ij}^\ast C_{jk}C_{ki})\sqrt{\lambda_i\lambda_j}
(F_1\ +\ F_2)\nonumber\\
&&-\ \Big(\ \mbox{Im}(C_{ij}^\ast C_{jk}^\ast C_{ki})\sqrt{\lambda_i\lambda_k}
\ +\ \mbox{Im}(C_{ij}C_{jk}^\ast C_{ki})\sqrt{\lambda_j\lambda_k}\ \Big)
(R\ +\ F_1\ -\ F_2) \Bigg],\ \
\eea 
where
\bea
R\ & =&\ \theta (M_H-m_{n_i}-m_{n_j})\ \frac{1}{2}\
\ln\left( \frac{t^+(\lambda_i,\lambda_j)-\lambda_k}
{t^-(\lambda_i,\lambda_j)-\lambda_k} \right)\nonumber\\
&&+\ \theta (M_W-m_{n_i}-m_{n_k})\ \frac{1}{2}\
\ln\left( \frac{\bar{t}^+(\lambda_i,\lambda_k)-\lambda_j}
{\bar{t}^-(\lambda_i,\lambda_k)-\lambda_j} \right)\nonumber\\
&&+\ \theta (M_W-m_{n_j}-m_{n_k})\ \frac{1}{2}\
\ln\left( \frac{\bar{t}^+(\lambda_j,\lambda_k)-\lambda_i}
{\bar{t}^-(\lambda_j,\lambda_k)-\lambda_i} \right).
\eea 
The functions $F_1$, $F_2$, $t^\pm$, and $\bar{t}^\pm$ emerge from
from Eqs.~(21)--(25) by substituting
 $\lambda_W\to \lambda_Z$ and $\lambda_{l_k} \to \lambda_k$.
{}From Table 1 we see
that $A_{CP}^{(Z)}$ can be of the order of $10\%$.
An effect of this size should be observable at the next $e^+e^-$ linear
collider.

In conclusion, we have seen that heavy Majorana neutrinos represent
an attractive alternative to complicated multi-Higgs-boson
scenarios~\cite{CPhig} when it comes to accounting for possible
$CP$-violating phenomena in the decays of Higgs
bosons into $t\bar{t}$, $W^+W^-$, and $ZZ$ pairs.
The minimal extension of the $SM$ by right-handed neutrinos (assuming
the Majorana mass matrix, $m_M$, to be bare and, hence, the absence of Majoron
fields) can naturally describe sizable $CP$-violating effects,
of the order of $10\%$, at high-energy $e^+e^-$ linear colliders.
Although the $CP$ asymmetries $A_{CP}^{(t)}$,
$A_{CP}^{(W)}$, and $A_{CP}^{(Z)}$ may not be directly accessible
by experiment, the $CP$-violating signals originating from such Higgs-boson
decays will, however, be transcribed to the decay products of the top-quark,
$W$-, and $Z$-boson pairs.
More realistic $CP$-odd projectors may be constructed
by considering, for instance, angular-momentum distributions or energy
asymmetries of the produced charged leptons and jets~\cite{Peskin,CH,BAK}.


\vspace*{-0.5cm}

\begin{thebibliography}{99}

\bibitem{CPhig} J.F.~Donoghue and G.~Valencia,
{\em Phys.~Rev.~Lett.}~{\bf 58} (1987) 451; {\bf E60} (1988) 243;\\
M.~Nowakowski and A.~Pilaftsis, {\em Mod.~Phys.~Lett.}~{\bf A6} (1991) 1933;\\
G.~Eilam, J.L.~Hewett and A.~Soni, {\em Phys.~Rev.~Lett.} {\bf 67}
(1991) 1979; {\bf 68} (1992) 2103 (Comment); R.~Cruz,
B.~Grzadkowski and J.F.~Gunion, {\em Phys.~Lett.}~{\bf B289} (1992) 440;
D.~Atwood, G.~Eilam and A.~Soni, {\em Phys.~Rev.~Lett.}~{\bf 71}
(1993) 492.

\bibitem{Peskin} C.R.~Schmidt and M.E.~Peskin,
{\em Phys.~Rev.~Lett.}~{\bf 69} (1992) 410.

\bibitem{CH} D.~Chang and W.-Y.~Keung, {\em Phys.~Lett.}~{\bf B305} (1993) 261.

\bibitem{PN} A.~Pilaftsis and M.~Nowakowski, Mainz University Report No.\
MZ-TH/92-56, {\em Int.~J.~Mod. Phys.~A} (to appear).

\bibitem{IKP} A.\ Ilakovac, B.A.\ Kniehl and A.\ Pilaftsis,
UW-Madison Report No.\ MAD/PH/787, {\em Phys.\ Lett.\ B } (to appear).

\bibitem{ZPC} A.~Pilaftsis, {\em Z.~Phys.}~{\bf C55} (1992) 275.

\bibitem{KPS}
J.G.~K\"orner, A.~Pilaftsis and K.~Schilcher, {\em Phys.~Rev.}~{\bf
D47} (1993) 1080.

\bibitem{LL} P.~Langacker and D.~London, {\em Phys.~Rev.}~{\bf D38} (1988) 886.

\bibitem{PH} C.T.~Hill and E.A.~Paschos, {\em Phys.~Lett.}~{\bf B241}
(1990) 96;
C.T.~Hill, M.A.~Luty and E.A.~Paschos, {\em Phys.~Rev.}~{\bf
D43} (1991) 3011.

\bibitem{BAK} V.~Barger, K.~Cheung, A.~Djouadi, B.A.~Kniehl and
P.M.~Zerwas, DESY Report No.\ 93-064, {\em Phys.~Rev.~D} (to appear);
B.A.~Kniehl, DESY Report No.\ 93-088, {\em Proceedings of the
Workshop on Physics and Experiments with Linear $e^+e^-$ Colliders},
Waikoloa, Hawaii, April 26--30, 1993 (to appear).

\end{thebibliography}
\end{document}